\RequirePackage{snapshot}
\documentclass[5p,preprint,authoryear,number]{elsarticle}

\usepackage{graphicx}
\usepackage{hyperref}
\usepackage{listings}
\usepackage{amsmath}
\usepackage{here}

\lstdefinestyle{pseudobase}{
	language=C++,
	emptylines=1,
	breaklines=true,
	basicstyle=\color{black}\ttfamily,
	keywordstyle=\color{blue}\ttfamily,
	commentstyle=\color{green!50!black}\ttfamily,
	morecomment=[l][\color{magenta}]{\#},
	moredelim=**[is][\color{blue!50!green}]{@}{@},
	deletekeywords={and,for}
}
\lstdefinestyle{numbered}{
	numbers=left,
	stepnumber=2,
	firstnumber=1,
	numberfirstline=false
}

\DeclareMathOperator{\erf}{erf}

\journal{Astronomy \& Computing}

\newcommand{\quotes}[1]{``#1"}

\begin{document}
	
\begin{frontmatter}
	\title{WVTICs - SPH initial conditions for everyone}
	\author[usm,excluster,mpe]{Arth, A.}
	\ead{arth@usm.uni-muenchen.de}
	\author[led]{Donnert, J.}
	\ead[url]{https://github.com/jdonnert/WVTICs}
	\author[usm,mpa]{Steinwandel, U. P.}
	\author[usm]{B\"oss, L.}
	\author[mpa]{Halbesma, T.}
	\author[konstanz]{P\"utz, M.}
	\author[usm,excluster]{Hubber, D.}
	\author[usm,mpa]{Dolag, K.}
	\address[usm]{University Observatory Munich, Ludwig-Maximilians-University Munich, Scheinerstr.1, D-81679 Munich, Germany}
	\address[mpe]{Max-Planck-Institute for Extraterrestrial Physics, Giessenbachstrasse 1, 85748 Garching, Germany}
	\address[led]{Leiden Observatory, PO Box 9513, NL-2300 RA Leiden, The Netherlands}
	\address[mpa]{Max Planck Institute for Astrophysics, Karl-Schwarzschild-Str. 1, D-85741 Garching, Germany}
	\address[konstanz]{Universit{\"a}t Konstanz, Fachbereich f{\"u}r Physik, 78457 Konstanz, Germany}
	\address[excluster]{Excellence Cluster Universe, Boltzmannstr. 2, D-85748 Garching, Germany}
	
	\begin{abstract}
		We present a novel and fast application to generate glass-like initial conditions for Lagrangian hydrodynamic schemes (e.g. Smoothed Particle Hydrodynamics (SPH)) following arbitrary density models based on weighted Voronoi tessellations and combine it with improved initial configurations and an additional particle reshuffling scheme. We show our application's ability to sample different kinds of density features and to converge properly towards the given model density as well as a glass-like particle configuration. We analyse convergence with iterations as well as with varying particle number. Additionally, we demonstrate the versatility of the implemented algorithms by providing an extensive test suite for standard (magneto-) hydrodynamic test cases as well as a few common astrophysical applications. We indicate the potential to bridge further between observational astronomy and simulations as well as applicability to other fields of science by advanced features such as describing a density model using gridded data for exampling from an image file instead of an analytic model.
	\end{abstract}
	
\end{frontmatter}

\section{Introduction}
	\label{sec:intro}

	With rising computing power, simulations have become an integral part of modern astrophysics over the last few decades. In the past years people have traversed more and more from theoretical calculations with pen and paper to complex computer driven computations. These range from high precision simulations of idealised systems on various scales, as for example \cite{Bonafede2011,Bate2013,Pakmor2013,Steinwandel2019}, to large cosmological boxes which model a significant portion of the visible universe \citep[e.g.][]{Hirschmann2014, Vogelsberger2014, Schaye2015}. Several numerical techniques have been developed, refined and compared in order to improve the tools to widen our understanding of the physical processes which take place in the universe.\\
	What they all have in common is the prerequisite of some sort of initial conditions for their simulations. These may be physically motivated as in the examples given above or simply pose the means to test the behaviour of one's code with an analytically understood problem. Not only a proper definition of the physical key quantities in these initial conditions is important, but also their numerical representation. Often simulations analyse physical processes in the highly non-linear regime, where noise produced from badly sampled initial conditions or even the round off error introduced by finite accuracy floating point numbers becomes important (e.g. \cite{Liska2003}).\\
	Depending on the numerical scheme, obtaining optimal initial conditions can be a nearly trivial task or rather complicated. The former being the case in finite difference and finite volume discretisation schemes and the latter, as in the focus of this paper, for  mass discretisation schemes such as Smoothed Particle Hydrodynamics (SPH) \citep{Lucy1977,Gingold1977} or Meshless Finite Mass / Volume \citep{Gaburov2011,Hopkins2015a}. In this paper, we present an open source code which can generate relaxed initial conditions for SPH given any arbitrary physical description of initial conditions based on the work of \cite{Donnert2014,Donnert2017}. We start by defining the actual task in section \ref{sec:icgenoverview}, followed by a description of the presented algorithms in section \ref{sec:codedescription}. In section \ref{sec:testproblems} we show the performance of our implementation measured by way of several test problems and finally present a few applications in section \ref{sec:application} before closing with describing the actual usage of the application in section \ref{sec:usage}.

\section{Overview IC generation}
	\label{sec:icgenoverview}
	
	\subsection{Requirements and degrees of freedom}
		\label{sec:icrequirements}
	
		Initial conditions contain a complete physical description of all physical quantities as a function of position inside the simulation volume, such as the density $\rho \left( \vec{r} \right)$, internal energy $u \left( \vec{r} \right)$ and the velocity $\vec{v} \left( \vec{r} \right)$. This mathematical description can be translated into a volume discretisation by volume averaging the functions, or in case of finite difference schemes, evaluating the functions at the discretisation points. For a mass discretisation scheme this step is not straight forward, because mass sampling particles have to be placed in a way that the physical quantities calculated from the ensemble are reproduced with minimal error. To this end, obtaining the SPH smoothing length \& density already requires solving the continuity equation \citep{Price2012}. The other quantities then fall into place and need not to be discussed further.\\

		In the SPH formalism the density is calculated as the kernel weighted sum over the nearest neighbours:
		\begin{equation}
			\rho_i = \sum\limits_j^{\#\mathrm{NGBs}} m_j W \left( \left| \vec{r}_i - \vec{r}_j \right|, h_{i} \right)
		\end{equation}
		The kernel function depends on the distance between both particles $i$ and $j$ and the smoothing length $h_i$. As standard kernel in our implementation and all tests presented in this paper we chose the Wendland C6 kernel function with 295 neighbours in three dimensions, following \cite{Dehnen2012}.
		Since this is the only SPH specific type of equation we need for this paper, we refer the interested reader to \cite{Price2012} for a recent review of the SPH method.\\
		To reach a certain model density one can choose to use either constant or variable mass particles. While the latter makes it possible to choose a universally ideal particle placement by adjusting the masses to fit the density, it is very disadvantageous. A significant contrast in particle masses leads to either a large variance in the amount of neighbour inside a kernel and to high computational cost in low density regions or even to numerical instabilities. \cite{Monaghan2006} have shown, that one can not reach a low energy state with variable mass SPH particles with the classical SPH method.\footnote{There exist modern hybrid methods for which this is not necessarily the case, like the MFV method presented in \cite{Hopkins2015a}.} Therefore, we use constant mass SPH particles.
	
	\subsection{Particle placement}
		\label{sec:placementmethods}
		
		\begin{figure}
			\centering
			\includegraphics[width=\columnwidth]{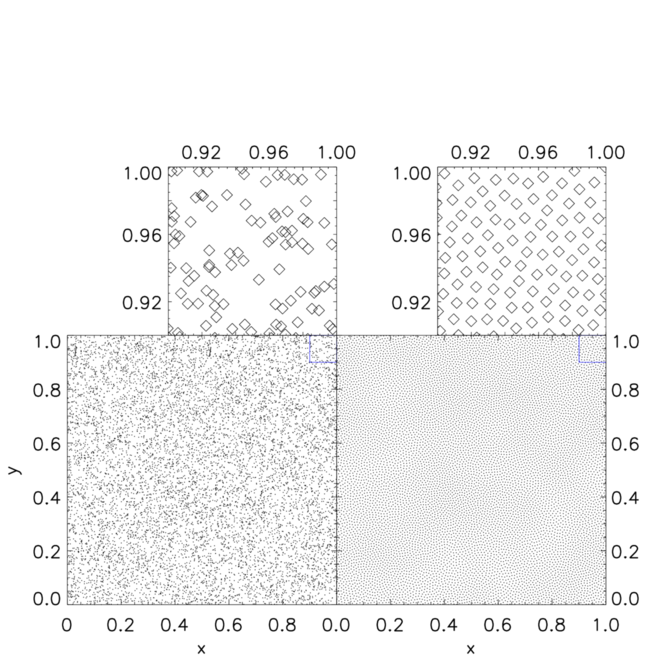}
			\caption{2D particle placement for a constant density model with $10^4$ particles for a random distribution (left) and a glass distribution (right). The upper panels show a zoom on each particular distribution in the spatial region between $x>0.9$ and $y>0.9$. While the random particle distribution leads to local clustering of the particles the glass distribution shows a regular distribution of the particles.}
			\label{fig:constant2Dparticles} 
		\end{figure}
		Even in the simplest case of a constant density several common approaches exist for particle placement. These include for example random positions, lattice configurations and so called glass files.\\
		The former is easy to implement, but the poison noise translates directly into a large relative error in the density estimate.\\
		In the left panel of Fig. \ref{fig:constant2Dparticles} we present a two dimensional, random particle distribution, which reveals an irregular structure full of holes and clumps. Furthermore, particles are more likely to be closer to each other than the mean particle distance, than further away, which introduces an additional anisotropy onto the density noise.\\
		Lattice configurations on the other hand come in different flavours, for example based on a cubic unit cell (cP, bcc or fcc) or in hexagonal configuration (hcp, \cite{Kotarba2011}). The advantage of these configurations is their regularity which results in an easy equilibrium configuration, i.e. a global minimal energy configuration. However, effects known from grid codes are introduced, like the introduction of principle axes which may lead to orientation-dependent results.\\
		Finally, in a glass file particles are distributed in a similar fashion as molecules in a physical glass. It defines a quasi hydrostatic equilibrium condition for the system and therefore poses (at least a local) a low energy state. Particles are distributed non-regularly, however, with a very confined distribution of inter particle distances which helps to reduce noise \citep{White1994}. This can be seen in the right panel of Fig. \ref{fig:constant2Dparticles}.
		
	\subsection{Variable density modifications}
		\label{sec:variabledensity}
		
		The initial random sampling can be easily adjusted to follow a different density distribution than the uniform one. We will later exploit this in section \ref{sec:initialstate}. However, the sampling error remains. Therefore, we rule out this method if it is used without additional effort to reduce the sampling error.\footnote{Random sampling still poses as a valid starting state for our approach.} When considering lattices and glass files, we remark that these can not be easily translated to a variable density. The homogeneous setup has to be stretched and compressed to follow the underlying density model. This process is not straight forward and it results in a loss of regularity for the lattice, while the glass transforms to a less relaxed ensemble. Nevertheless, stretching of uniform distributions is still commonly applied in the community \citep[e.g.][]{Price2005}. We refer to \cite{Diehl2012} for a comparison of different approaches.\\
		\citet{Hubber2016,Hubber2017} present the novel SPH/N-body code Gandalf that uses an alternative method for the generation of initial conditions. This method relaxes a set of particles towards a density model and a glass distribution simultaneously by applying two forces between particle pairs, one for each target property. The algorithm we present in this paper works similarly at it's core and we proceed by describing our approach in the next section.

\section{Code description}
	\label{sec:codedescription}
	
	Our implementation is based on a Weighted Voronoi Tesselation (WVT) as presented in \cite{Diehl2012} and aims to produce relaxed, low energy glass files for any arbitrary given physical model. In the following subsections we will first describe the core parts of the algorithm, then clarify issues that typically arise and finally present a quick study on how to improve the convergence by choosing the best possible starting configuration.

    \subsection{WVT relaxation}
	    \label{sec:relaxation}
	    
\begin{lstlisting}[style=pseudobase,basicstyle=\small,frame=single,float,floatplacement=H,label=lst:pseudocode,caption=Algorithm in pseudo code]
Initial particle placement
Do Until converged:
  Calculate Density, Smoothing Lengths
  For each particle loop over neighbours:
    Distance vector between particle pair
    Weight with expected smoothing & kernel
    Push particles apart
  Check convergence criteria
Set other quantities at particle positions
\end{lstlisting}
	    We do not need to implement the actual Voronoi tessellation, although the original algorithm is based on a Voronoi Tesselation as shown by \cite{Diehl2012} (hence the name WVT). A short pseudo code describing the main routine is given in listing \ref{lst:pseudocode}. The relevant parts of the relaxation algorithm are wrapped inside a generic SPH code which calculates densities and smoothing lengths for the current particle configuration. In each relaxation step a neighbour sum in classical SPH fashion is executed and for each particle pair a force is calculated to push the particles apart. The force is weighted with both the target smoothing length and the distance of both particles in order to converge to the given density model as well as a glass-like structure. In addition, a scaling factor for the movement of particles is set depending on the mean inter-particle distance. The goal is to find the optimum for the magnitude of particle pushes between wasting computing time through small steps and the possibility of overshooting. The net particle displacement of particle $i$ due to particle $j$ is then given by
	    \begin{equation}
		   \Delta \vec{r}_i = \text{const} \cdot h_{ij}^m \cdot W \left( \left| \vec{r}_{ij} \right|, h_{ij}^m \right) \cdot \frac{\vec{r}_{ij}}{\left|\vec{r}_{ij}\right|}
	    \end{equation}
	    with the expected model smoothing length
	    \begin{equation}
		    h_i^m = h^m \left( \vec{r}_i \right) = \left( \frac{N_{{NGB}} \cdot m}{\frac{4}{3} \pi \rho^m \left( \vec{r}_i \right)} \right)^{1/3},
	    \end{equation}
	    which is normalised such that it equals to
	    \begin{equation}
		    h_i^m = \left( \frac{1}{\frac{4}{3} \pi} \cdot \frac{\rho^m \left( \vec{r}_i \right)}{\sum\limits_j \rho^m \left( \vec{r}_j \right)} \right)^{1/3}.
	    \end{equation}
	    Furthermore, it enters in symmetrised form depending on both particles $i$ and $j$:
	    \begin{equation}
		    h_{ij}^m = \frac{\left( h_i + h_j \right)}{2}.
	    \end{equation}
	    There are two main changes made compared to the algorithm proposed by \cite{Diehl2012}: First, instead of an arbitrary weighting function which declines with distance we employ the most natural choice and use the kernel function from the surrounding SPH code, following \cite{Donnert2017}. Second, we switch out the multiplicative factor of $h_i^m$ and use a symmetrised version of it. Consider an ideal distribution of particles that exactly follows the model density. The algorithm should then not move particles, meaning $\Delta \vec{r}_i = 0 ~ \forall i$. In a simplified picture the push from a neighbour on one side and one on the opposing end should cancel each other out. Although, that is not the case any more with the symmetrised version, the use case for our code is to start with an imperfect particle distribution and iterate towards the desired solution. Tests show that we never reach the exact model solution globally and that the symmetrised version gives much better convergence towards it, which results in lower density errors. Whether we employ the arithmetic or geometric mean does not make any notable difference, so we keep the arithmetic one.\\
	    Furthermore, the algorithm can in principle get stuck in a local energy minimum, because it is formulated locally, on neighbouring particle pairs. However, under the condition that the density structures are well resolved (meaning kernels are not large enough to smooth them out), convergence is generally not an issue. We present an analysis of the convergence in section \ref{sec:testproblems}.\\
        Optimal detection of convergence poses a challenging task as well. Typical criteria formulated on the average and the maximum density error have proven to waste computation time without the further improving the sampled density profile. Therefore, we employ a more pragmatic criterion and stop the relaxation as soon as the most particles are moved less than a small fraction of inter-particle distance. In addition, we put an upper limit onto the maximum number of iterations to further reduce computational cost.

    \subsection{Initial state}
	    \label{sec:initialstate}
	    
	    The rate of convergence is drastically influenced by the initial state of the particle distribution. Thus, we compare a few different approaches:
	    \begin{itemize}
	    	\item Uniform random distribution
	    	\item Von Neumann rejection sampling
	    	\item Statistical model along a space filling curve
	    \end{itemize}
		The simplest approach for an initial setup is just a random arrangement of particles with positions drawn from a uniform distribution. As already discussed in section \ref{sec:placementmethods} this itself does not pose a very good density estimate but can still serve as a valid starting point for relaxation. Due to the distance dependence in our algorithm the most crucial flaw of a random distribution, namely holes and clumps of particles, are dissolved quite rapidly.\\
		In cases where the density model varies clearly from a uniform density, one can improve the initial state by changing the random number distribution to the density function itself. This ensures that the particle density already follows the model much better without changing other properties. This is known as Von Neumann rejection sampling \citep{VonNeumann1951} and is basically a Monte Carlo integration method. While the rejection can sometimes be costly for extremely heterogeneous distributions, the computational effort is still negligible compared to the relaxation afterwards.\\
		Despite the Poisson noise due to particle placement, random distributions can still be useful when the noise is energetically unimportant compared to a physical process, e.g. in driven supersonic turbulence simulations.\\
		We experimented with a different approach, formulating a statistical particle placement model long a space filling curve (e.g. a Peano curve \citep{Peano1890}) which gives very similar results while being more complicated to formulate properly. Therefore, we do not go into further details.
		
	\subsection{Additional redistribution}
		\label{sec:redistribution}
		
\begin{lstlisting}[style=pseudobase,basicstyle=\small,frame=single,float,floatplacement=H,mathescape=true,label=lst:redistributionpseudo,caption=Additional redistribution in pseudo code]
Calculate max amount and probes
dependent on iteration

While redistributed particles < max amount
and probes < max probes

  Draw untouched random particle $i$
  increase probes until accepted for movement
  $\forall$ i tested increment probes
  
  Draw random particle $j$
  until accepted as destination
  
  Random position $x$ with $|x - x_j| < 0.3 h_j$
  Move $i$ to position $x$
  Mark $i$ as touched
\end{lstlisting}
		As we will show, the WVT algorithm converges asymptotically towards the given model. It generates a glass distribution of particles settled in a local minimum of the potential energy defined by the pairwise inter-particle force. Since the algorithm operates only locally using an SPH kernel weighting, a global force term can help to avoid local energetic minima and improve global convergence. To this end, we include the possibility for an additional global particle redistribution step in our code, which is carried out every few relaxation steps. \\
		The idea behind this approach is that particles from regions with $\rho > \rho_\mathrm{model}$ are taken and directly moved into regions with $\rho < \rho_\mathrm{model}$. We design an algorithm which is similar to the well known Metropolis algorithm \citep{Metropolis1953} with the relative density error serving as the energy analogue. On top of that we put an upper limit to the amount of particles sampled, so that less particles are redistributed the closer we come to the model solution. We present corresponding pseudo code in listing \ref{lst:redistributionpseudo}. We randomly select particles to be moved and particles which we can use as tracers of an under dense region. Then we place the respective particles into the close proximity of these regions by drawing a random position with maximum distance of 0.3 of the kernel support radius. To accept a particle $i$ for redistribution, we check if a random number $r_i$ fulfils
		\begin{equation}
		r_i \in [0,1]  < \erf \left( \frac{\rho_i - \rho^m \left( \vec{x}_i \right)}{\rho^m \left( \vec{x}_i \right)} \right).
		\end{equation}
		Similarly a particle $j$ is viable as destination if another random number $r_j$ fulfils
		\begin{equation}
		r_j \in [0,1]  < \frac{\rho^m \left( \vec{x}_j \right) - \rho_j}{\rho^m \left( \vec{x}_j \right)}.
		\end{equation}
		This ensures that particles with large density errors are in favour of being redistributed and that more particles are probed the closer the density comes to the model. We limit ourselves to a percentage of the particles in order to preserve the overall density profile and allow us to omit density calculations in between this process. Otherwise the computational cost would be too high for the method to be feasible. The maximum number of particles to be redistributed and to be probed is given by runtime parameters. In order to assist convergence we also let both percentages decay exponentially over time. For example we start at $1 \%$ of particles and decay down to $0.1 \%$ of particles until a certain iteration step. Between each redistribution we carry out several relaxation iterations to smooth out the distribution. At some point we stop the redistribution and only the relaxation step is carried out. The impact of this additional algorithm is investigated in section \ref{sec:redistributiontests}.
		
\section{Test Problems}
	\label{sec:testproblems}
	
	In this section we present algorithmic behaviour in typical use cases. While it is fairly straight forward to measure the accuracy of the resulting density distribution, the quality of the glass is more difficult to judge on only by using properties of the particle distribution. We start with a quick analysis of generated constant density distributions and then go one step further to our actual application of variable density models. All test cases are carried out in 3D with periodic boundaries if not mentioned otherwise. In order to save computational time we generally set up non-cubic boxes. We run all tests for a maximum of 1024 iterations but terminate in general earlier when reaching a quasi steady state. The additional redistribution of particles is switched off. We discuss the effect of the redistribution scheme later.

	\subsection{Constant density}
		\label{sec:constantdensitytests}
		
		\begin{figure}
			\centering
			\includegraphics[width=\columnwidth]{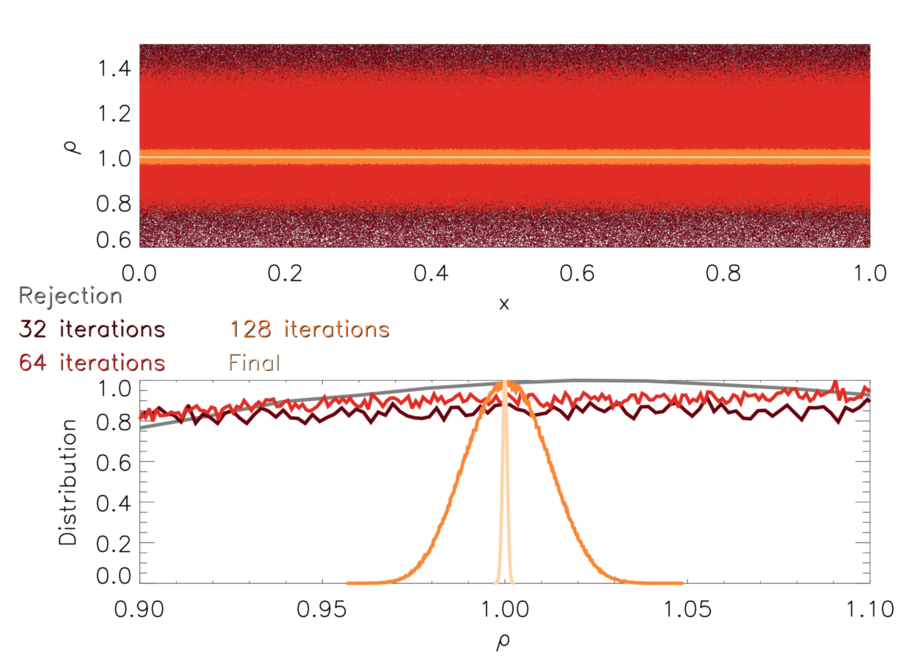}
			\caption{Evolution of the density for a 3D constant density model with $10^6$ particles for different number of iterations at the rejection step (grey), $32$ steps (dark red), $64$ steps (red), $128$ steps (orange) and the final step (light orange). The upper panel shows the density as a function of the distance. With increasing iterations the code evolves towards a model with the target density of $1$. In the bottom panel we show a mass weighted PDF to quantify the density distribution of the particles for different iteration steps. Further, this panel indicates the error that is introduced by the initial random distribution which is quickly reduced with increasing number of iterations.}
			\label{fig:constantdensity}
		\end{figure}
\begin{figure}
			\centering
			\includegraphics[width=\columnwidth]{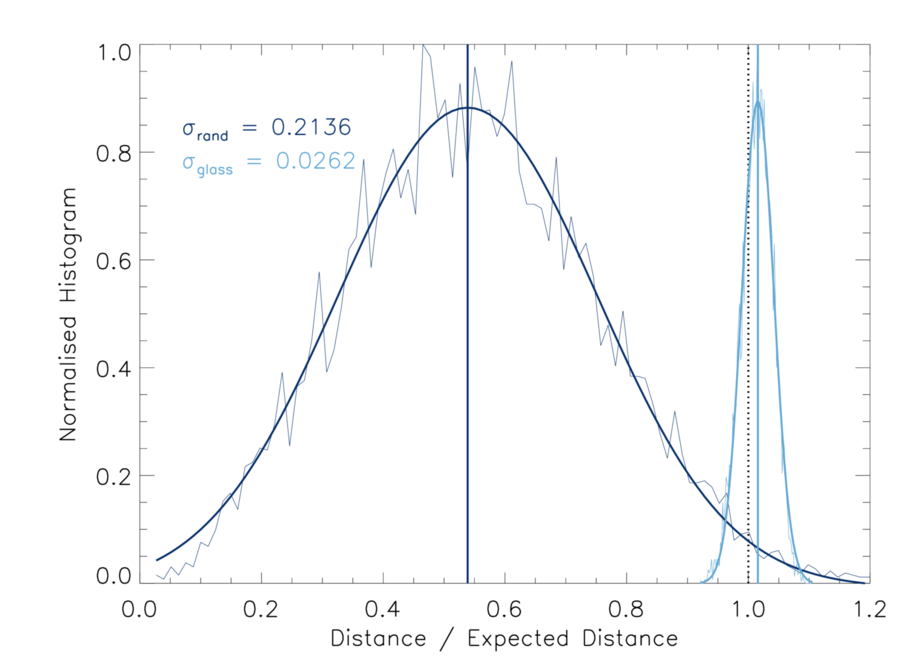}
			\caption{Histogram of nearest neighbour distances for a 3D constant density model for random placement and a glass distribution. The expected distance by which the x-axis is normalised is calculated by dividing the whole box volume up equally onto all particles.}
			\label{fig:nearestngbdistance}
		\end{figure}
\begin{figure}
			\centering
			\includegraphics[width=\columnwidth]{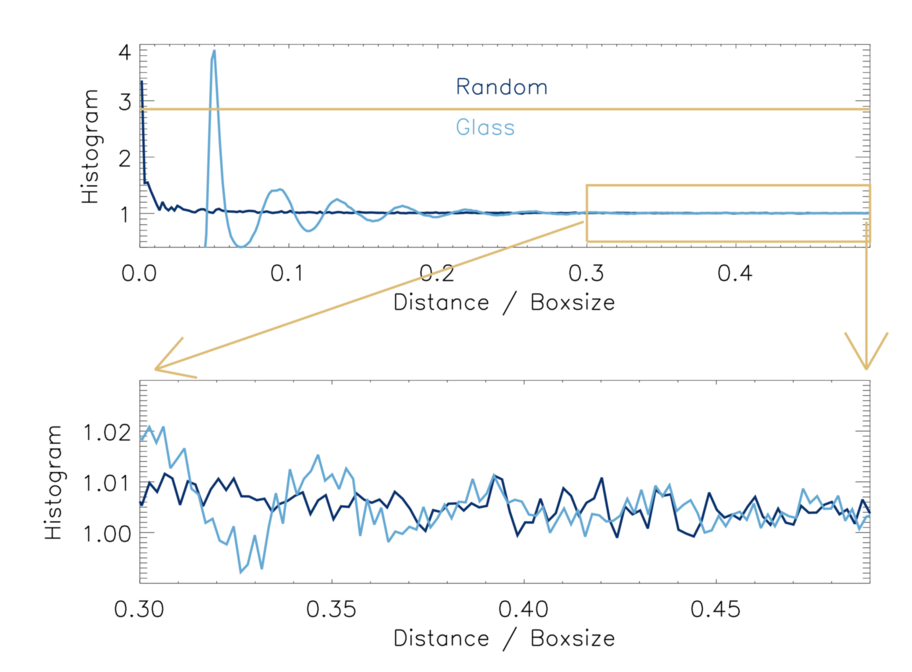}
			\caption{Auto correlation function for a 3D constant density model for an Lennart-Jones fluid model (dark blue) and a glass distribution (light blue) in the top panel. In combination with the Hansen-Verlet criterion (dark orange line) the auto correlation function can be used to measure the quality of the glass distribution that is obtained with our algorithm. In the bottom panel we show a zoom onto the large scales of our simulation domain where we find excellent agreement of our glass distribution with the theoretical model of a glass (below $1$ per cent deviation).}
			\label{fig:autocorrelation}
		\end{figure}

		We show how the density evolves for a constant density model during relaxation in Fig. \ref{fig:constantdensity}. The top plot shows the density values of all SPH particles after different number of relaxation steps while the lower plot presents the according distribution of density values zoomed onto a $\pm 10\%$ region of density values. Initially we sample the typical Poisson distribution, with noise of the order of $\pm 50\%$ around the target value, visible in the top plot of Fig. \ref{fig:constantdensity}, and quickly reduce to a reasonable agreement with the density model. One can clearly see the asymptotic behaviour of the convergence until the algorithm stops after $652$ iterations.\\
		To assess the quality of the particle distribution, an analysis similar to Fig. \ref{fig:constant2Dparticles} can reveal the emerging patterns only in an obscured way due to the integrated third dimension. Furthermore, it lacks the power of a proper quantitative analysis with a parameter to measure the quality. A better judgement can be made by looking at the distribution of nearest neighbour distances (Fig. \ref{fig:nearestngbdistance}) and the radial distance autocorrelation function\footnote{Or just \quotes{radial distribution function} in the field of molecular dynamics.} (Fig. \ref{fig:autocorrelation}).\\
		We compare these two measurements for a random distribution and a relaxed glass distribution for a 3D constant density model with only $10^4$ particles in a periodic box to prevent running into memory issues in our analysis script. We scale the distance by the expected distance calculated from a perfect distribution of the volume in the periodic box using
		\begin{equation}
		d = \left( V_\mathrm{p} \right)^{1.0/3.0} = \left( V / N_\mathrm{p} \right)^{1.0/3.0}
		\end{equation}
		with $V$ the volume of the whole box and $N_\mathrm{p}$ the particle number. Note that while this is the optimal distribution of volume onto all particles it does not present a feasible configuration. Even in best close packing of spheres, there is still about $22.2\%$ of the volume lost \citep{Muder1988}. This is reached by a proper hcp lattice and depicts the asymptotic target for a well relaxed glass distribution. The distributions in Fig. \ref{fig:nearestngbdistance} can be fitted by a Gaussian. The deviation of the mean value from the expected mean and the width of the fit can be taken as a measurement of quality of the distribution. Since particles tend to clump in a random distribution the whole distribution shifts to smaller distances. The width is dominated by the Poisson noise seen in the density distribution of Fig. \ref{fig:constantdensity}. While smaller distances are in principal desirable, this figure completely obscures that when particles clump some also spread too far apart from each other. Therefore, we have to not only consider the next neighbours but better analyse the whole distribution of particles using the autocorrelation function. 
		It gives us even more insight into how well our particle distribution resembles a physical glass. We bin the distances of all particle pairs and divide each resulting value by
		\begin{equation}
		\mathrm{Norm} = \frac{V_\mathrm{bin}  \left<\rho\right> \cdot N_\mathrm{p}}{m},
		\end{equation}
		with $V_\mathrm{bin}$ the bin's spherical shell volume, $\left<\rho\right>$ the mean density in the bin, $N_\mathrm{p}$ the particle number and $m$ the particle mass as presented in \cite{Frenkel2001}.\\
		We analyse the data by creating a histogram of the distribution (Fig. \ref{fig:autocorrelation}) which we do not further normalise or stretch in any way. The results from the different approaches can be compared to a Lennard-Jones fluid model as presented for example in \cite{Frenkel2001}. Two criteria of quality are the height of the first peak and the behaviour on large scales. First of all a glass is characterised by a main peak height of 2.85 according to the Hansen-Verlet criterion \citep{Hansen1969} which characterises the transition from a liquid to a frozen state, demonstrating that our glass is indeed well relaxed. Also the peak's location is important. Since the first peak is mainly dominated by next neighbours we did this comparison already in Fig. \ref{fig:nearestngbdistance}. While the random sample damps down pretty quickly to a fairly constant distribution of values, we see dampened wave-like behaviour for our glass distribution around a central value of about one, as expected. The zoom onto large scales in the lower panel of Fig. \ref{fig:autocorrelation} especially shows, that this extends out properly until the maximum distances given by half of the box' size, showing that the glass distribution we produce is of high quality.\\
	
	\subsection{Variable density}
		\label{sec:variabledensitytests}
		
		\begin{figure}
			\centering
			\includegraphics[width=\columnwidth]{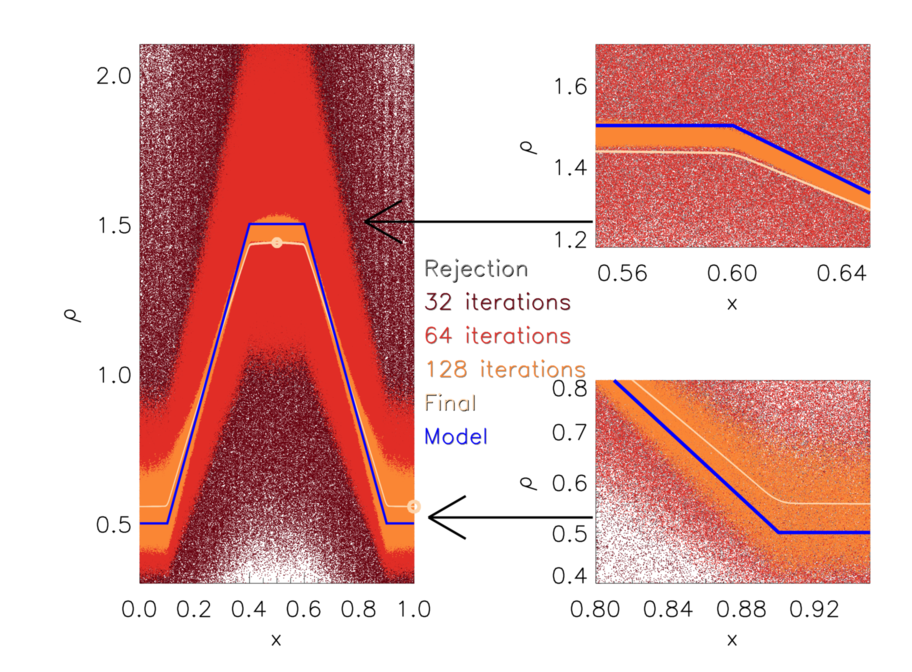}
			\caption{Evolution of a plateau density function with steep gradients on the edges for a particle number of $10^6$. In the left panel we show density plateau for $32$ iterations (dark red), $64$ (red), $128$ (orange) and the final iteration step (light orange) and compare it to the analytic density profile (blue). In the two panels on the left we show zooms onto the transitions from constant densities to the linear gradient for the low density region (bottom) and the high density region (top). This sharp transition regions from constant densities to steep density gradients are extremely difficult to resolve. However, the algorithm manages this with an accuracy of around $1$ per cent.}
			\label{fig:1E6tophatrejection}
		\end{figure}
		Transitioning to variable density models, the next test consists of two plateaus connected with linear slopes. Here we examine the behaviour of our implementation with respect to linear gradients and kinks in the density function. In Fig. \ref{fig:1E6tophatrejection} we present the evolution of the resulting density profile along with the analytic model. In addition to the overall profile, we also show zooms onto one upper and one lower kink of the distribution. The analytic model is given by
		\begin{equation}
		\rho(x) = \left\{\begin{array}{ll}
		0.5, & \text{for } x < 0.1\\
		0.5 + (x-0.1)/0.3 & \text{for } 0.1 < x < 0.4\\
		1.5 & \text{for } 0.4 < x < 0.6\\
		1.5 - (x-0.6)/0.3 & \text{for } 0.6 < x < 0.9\\
		0.5 & \text{for } x > 0.9.
		\end{array}\right.
		\end{equation}
		Again the scatter reduces quite rapidly and we converge towards the model density. As the scatter decreases, the density flattens marginally with respect to the actual maxima and minima of the given density function. Therefore, a low energy state and following the proper density model seem to be slightly exclusive. Since the extremal values are not fully resolved, the gradients are marginally shallower than imposed. Also the constant density parts are not totally flat but show a slight curvature which is expected due to the influence of their surroundings. SPH can not produce any sharp edges by construction, only up to the resolution level given by the kernel size. We over plot two circles in both the high and low density region in order to give an idea of the size of the respective kernels in these regions to show that the model is well resolved with $10^6$ particles. In the end, after 1024 iterations steps, we reach a steady solution with a very low degree of scatter which deviates from the analytic solution by a few percent and shows slight curvature in the plateau and smoothed out kinks which, however, sit at the expected $x$-values.\\
		\begin{figure}
			\centering
			\includegraphics[width=\columnwidth]{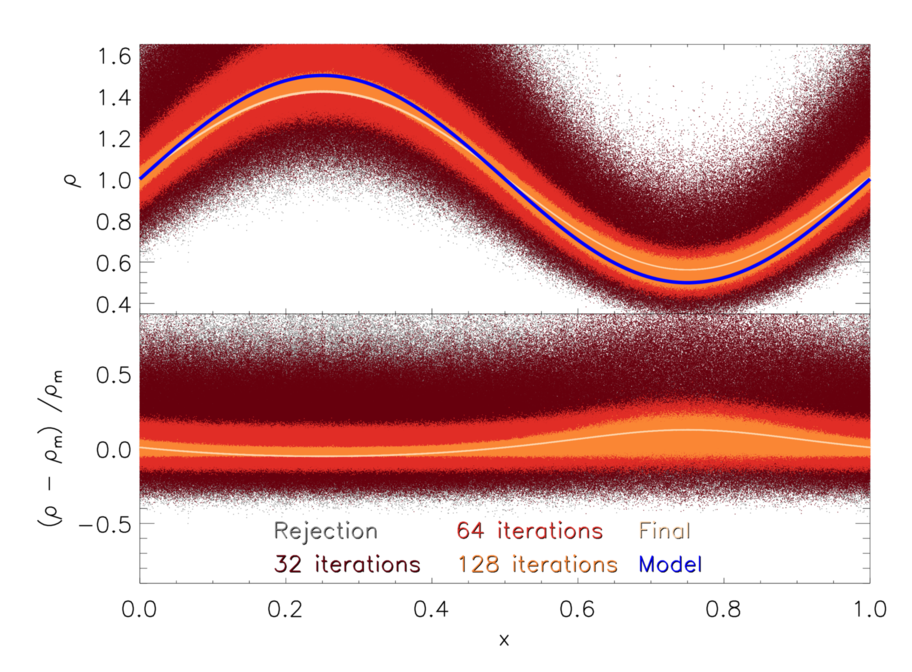}
			\caption{Evolution of a one dimensional sine wave perturbation as initial condition for a three dimensional simulation for $10^{6}$ SPH-particles The top panel shows the evolution of the absolute density value as for the von Neumann-rejection sampling (grey), after $32$ iterations (dark red), after $64$ iterations (red), after $128$ iterations (yellow) and at the final iteration (gold). Additional, we show the underlying analytic model density in blue. We find very good agreement with the model density up to a few per cent. In the bottom panel we show the relative density error for the different number of iterations and find very good agreement at the final iteration. However, we note that we find a slight bump at the density minimum. This can be explained by the larger kernel noise on the per cent level in the low density regime compared to the high density regime. }
			\label{fig:1E6sinewaverejection}
		\end{figure}
		Next, we look at a smoother density model to investigate the error in maxima / minima further without the introduction of sharp edges or linear gradients. We set the density model according to
		\begin{equation}
		\rho(x) = 1 + \frac{1}{2} \cdot \sin (2 \pi x)
		\end{equation}
		and plot the resulting density evolution in Fig. \ref{fig:1E6sinewaverejection}. Due to the smooth nature of this test it already converges after 865 iterations. When reducing the scatter to a very low degree the density function again slightly deviates from the model at the maxima and minima, proving that this is not an effect of non-smooth density definitions. The lower panel shows the resulting relative density error compared to the imposed model for all plotted steps. Although slightly asymmetric the error being of the order of a few percent is definitely satisfactory.\\
		\begin{figure}
			\centering
			\includegraphics[width=\columnwidth]{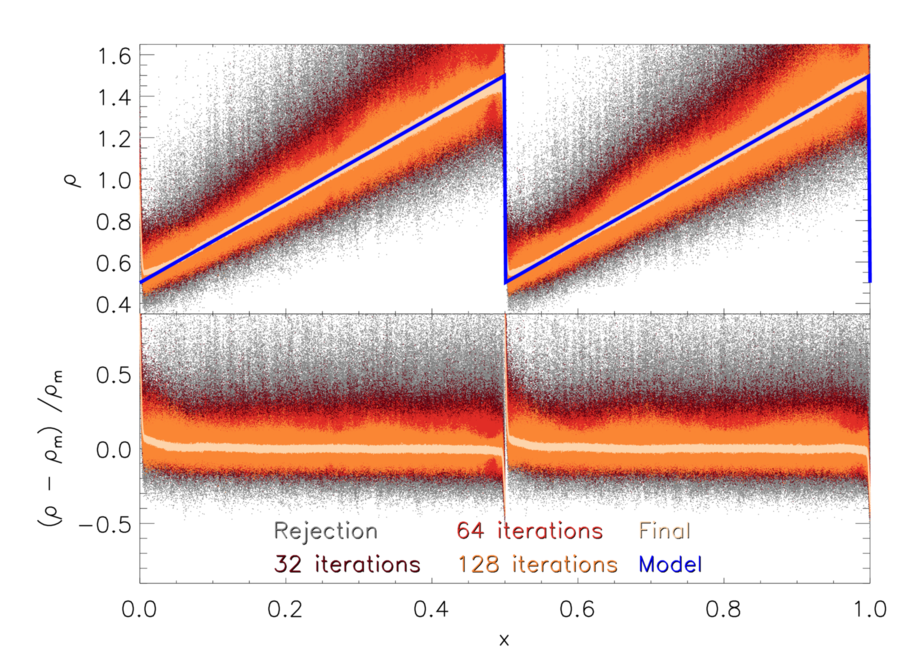}
			\caption{Evolution of a one dimensional sawtooth as an initial condition for a three dimensional simulation for $10^{6}$ SPH-particles The top panel shows the evolution of the absolute density value as for the von Neumann-rejection sampling (grey), after $32$ iterations (dark red), after $64$ iterations (red), after $128$ iterations (yellow) and at the final iteration (gold). Additional, we show the underlying analytic model density in blue. We find very good agreement with the model density up to a few per cent. In the bottom panel we show the relative density error for the different number of iterations and find very good agreement at the final iteration.}
			\label{fig:1E6sawtoothrejection}
		\end{figure}
		Finally we  investigate our algorithm's behaviour when confronted with sudden density jumps. We set the target density to
		\begin{equation}
		\rho(x) = \frac{1}{2} + 2 \cdot \left( x \mod \frac{1}{2} \right)
		\end{equation}
		and plot the result in Fig. \ref{fig:1E6sawtoothrejection}. While we can observe slightly more scatter here than in the previous tests, indicating that the final state (1009 iterations) is probably less converged due to the presence of the prominent density discontinuity, the jump itself is well resolved. Already after 32 iterations hardly any particles populate intermediate densities between the high and the low state. Also the resulting density error is in the same regime as before. We conclude that our implementation is able to reproduce all kinds of density models within the boundaries of SPH itself and the imposed resolution. Only density models which contain regions with $\rho = 0$ are challenging to model and require some fine tuning in our current neighbour finding routine.\\
		\begin{figure}
			\centering
			\includegraphics[width=\columnwidth]{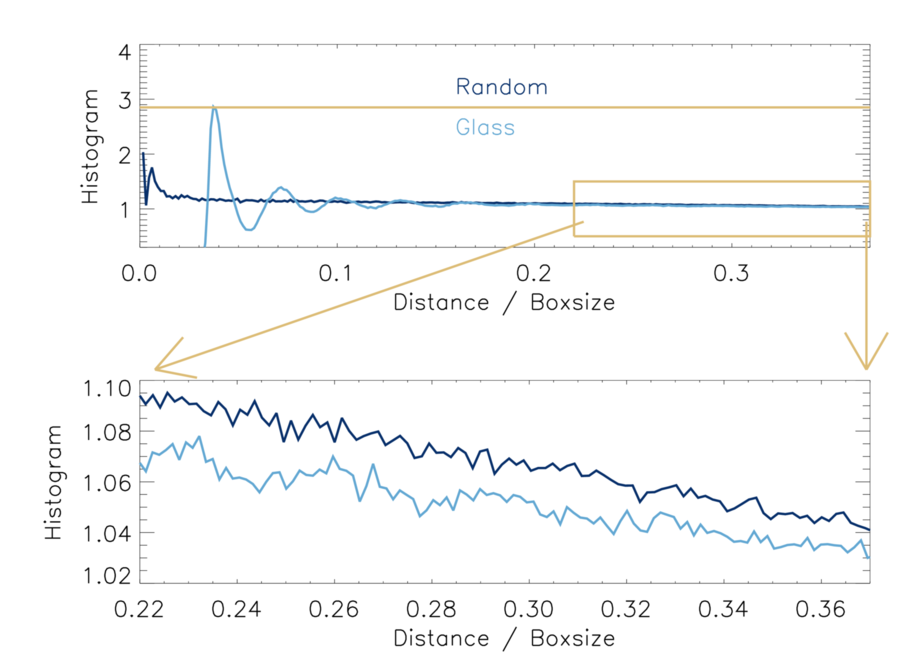}
			\caption{Auto correlation function for the sine wave density distribution with a particle number of $10^4$. The upper panel shows the auto correlation function as a function of distance for a random distribution (dark blue) and a glass distribution (light blue). The Hansen-Verlet criterium that marks the transition from a fluid to a solid (glass-like) state is indicated by the horizontal orange line.}
			\label{fig:autocorrelationsine}
		\end{figure}
		Considering the quality of the produced glasses,  we plot again the autocorrelation function for $10^4$ particles distributed according to the sine wave model density in Fig. \ref{fig:autocorrelationsine}. While the mean density is still at the same level as in section \ref{sec:constantdensitytests}, this introduces higher and lower density regions. High density regions contain more particles than the average. This leads to a shift in the distribution to smaller distances, especially noticeable at the main peak. In addition, we see a more strongly damped wave pattern compared to the previous tests. Furthermore, we notice that the distribution declines slightly when going to large scales. We have to truncate this plot at smaller scales since we divide the distances by the box size in x direction while the y and z direction are slightly smaller here to save computing time.\\
		We conclude that our 3D particle distributions actually form satisfactory glass configurations.
	
	\subsection{Including Particle Redistribution}
		\label{sec:redistributiontests}
		
		Now we investigate the effects of our particle redistribution scheme (section \ref{sec:redistribution}). We start to probe ten percent and to redistribute at maximum one percent of all particles. We carry out this algorithm every 5 iterations until we reach the 512th iteration step and let the percentages decay exponentially to go down a factor of ten until the end. We carry out all three presented variable density tests with this setup and compare to the results presented in section \ref{sec:variabledensitytests}.\\
		The sine wave density shows a bit larger scatter until the redistribution is switched off and the end results improves only slightly. However, the algorithm now requires about twice the iterations as before. While this can be mitigated by adjusting parameters, the improvement due to redistribution seems hardly worth it. The sawtooth setup shows minimal improvement, only the discontinuity becomes a bit sharper and less round.
		\begin{figure}
			\centering
			\includegraphics[width=\columnwidth]{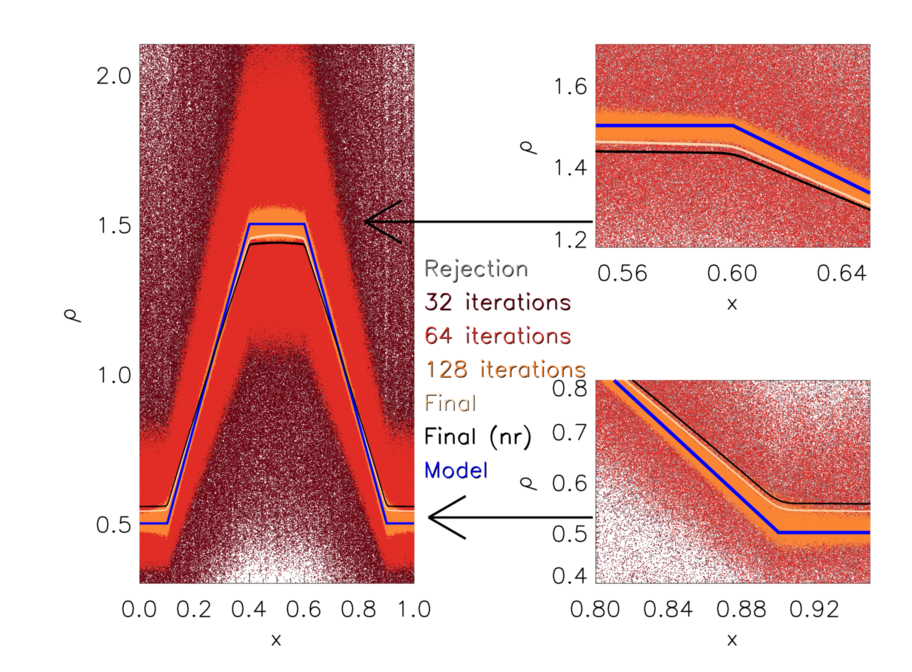}
			\caption{We show a comparison with and without our particle redistribution scheme for the top hat density distribution with $10^6$ particles. On the left hand side we show the total density distribution for the top hat density distribution with the rejection sampling (grey) for $32$ iterations (dark red), $64$ iterations (red), $128$ iterations (orange) and the final iteration step (gold) with the particle redistribution scheme. We compare that to the results that we obtain without the particle redistribution scheme (black) and the analytic density profile (blue). We find slight improvement by using the particle redistribution scheme. On the right hand side we show zooms onto the edges of the density distribution. We find an improvement of roughly two per cent by using the redistribution scheme compared to the results that we obtain without the redistribution scheme.}
			\label{fig:redistcomp}
		\end{figure}
		The plateau density setup shows the biggest effect, we plot the result in Fig. \ref{fig:redistcomp}. The colours indicate the same iteration steps as before in Fig. \ref{fig:1E6tophatrejection} for better comparison and we include the final result from before as a black line. As can be clearly seen, the distribution settles closer to the actual model density. In this case the amount of iterations the algorithm  takes to converge is actually about the same in both cases. However, due to the additional disturbance from redistribution, the density plateau comes out less straight than before. 
		
	    \begin{figure}
	    	\centering
	    	\includegraphics[width=\columnwidth]{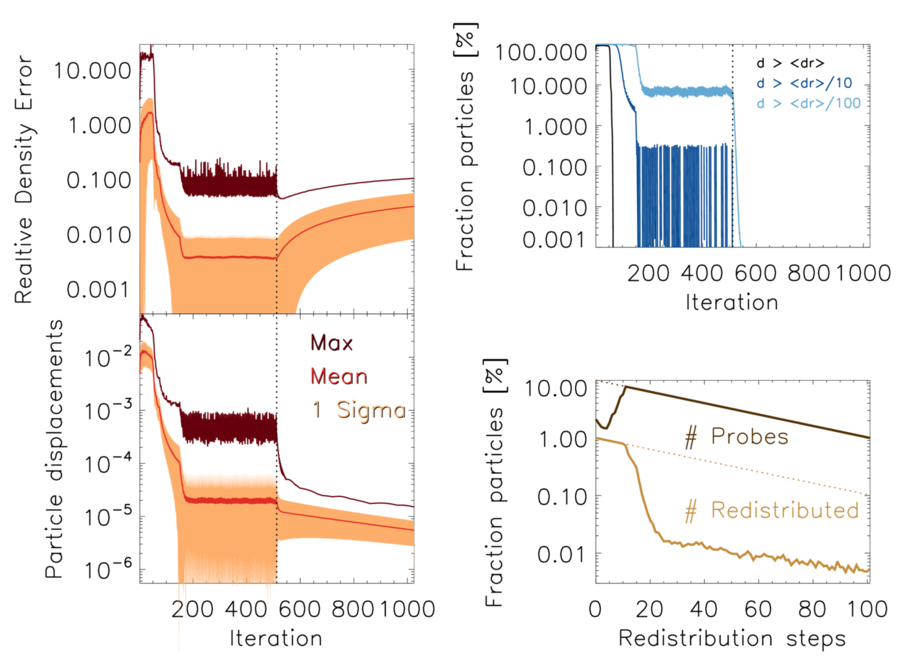}
	    	\caption{Plateau \& gradients test with $10^6$ particles. Top left panel: Min, mean and max density errors; Top right panel: Fractions of particles moved farther than 1, 0.1 and 0.01 of the local mean particle separation; Bottom left panel: Min, mean and max displacements calculated; Bottom right panel: Particles probed and actually redistributed; All plotted over the course of all 1024 iterations.}
	    	\label{fig:errorshiftiterations}
	    \end{figure}
	    To investigate the iteration process further, we sample the density plateau and linear gradient tests with $10^6$ particles and particle redistribution. Each iteration we write out on the global density error, the movement of all particles in the last step and the degree of particle displacements. We plot the results in Fig. \ref{fig:errorshiftiterations}. In the beginning we observe large particle displacements due to large density errors. These damp down rather quickly. At about 100 iterations, the maximum density error has gone down to a few percent while the mean error is even about a factor 10 smaller.\\
	    As the errors decline, a periodic pattern emerges which is the imprint of our particle redistribution scheme. As it removes particles from local over-densities and puts them into under-dense regions, injecting potential energy in the particle distribution. The algorithm damps this perturbation very quickly. Nevertheless, redistribution counteracts convergence and we can not see the improvements in the final density correctness at this resolution. The amount of particles marked for redistribution quickly reaches the allowed maximum and the amount of particles actually touched declines drastically down to $10^{-4}$ and then follows the imposed decay. We mark where this part of the code is shut off in the plots by a dotted vertical line. We do not improve on the density error any more but rather sacrifice a bit of density correctness for a better glass distribution and therefore less scatter in the density. The net particle kicks decrease as the distribution relaxes and movement damps down quickly. One can clearly see, that the net particle movement is well suited for an abortion criterion since it experiences the quickest change and directly indicates a steady state. \\
		We conclude that the improvement due to particle redistribution is highly problem dependent and not always worth the extra computational effort.
	    
	\subsection{Convergence with Particle Number}
		\label{sec:convergencevariablenpart} 

	    \begin{figure}
	    	\centering
	    	\includegraphics[width=\columnwidth]{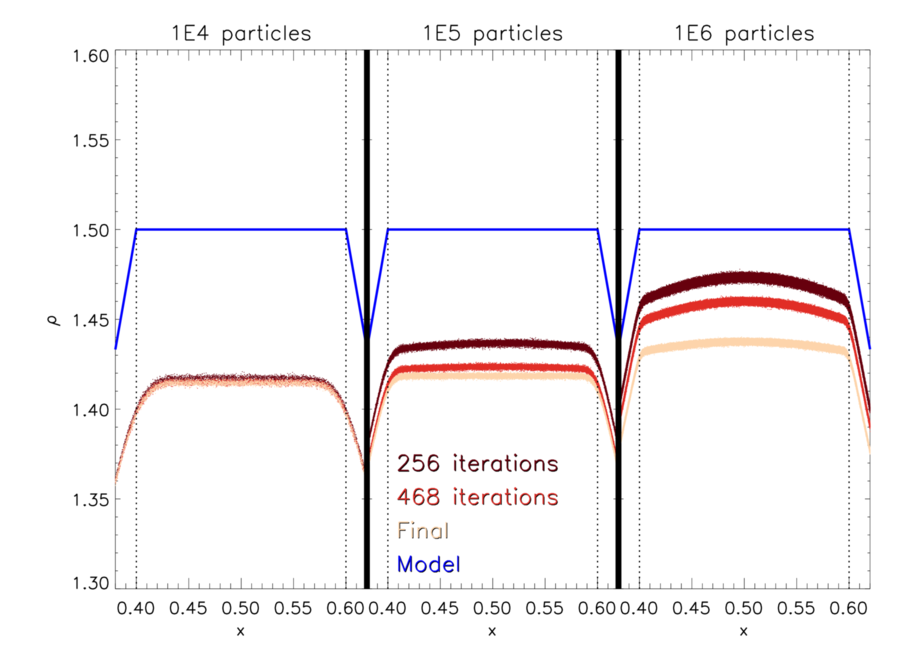}
	    	\caption{Resolution study of the plateau test for $10^4$ (left), $10^5$ (middle) and $10^6$ particles (right) for $256$ iterations (dark red), $468$ iterations (red) and the final iteration step (gold). We show the analytic model density in blue. With increasing number of iterations the code comes closer to the model density on the plateau. For $10^4$ particles at the final iteration we find that the code reproduces the density plateau within seven per cent. With increasing particle number higher accuracy can be obtained for $10^5$ particles we find a final deviation of the model density and the sampled distribution of four per cent. This number further improves to $1.5$ per cent at $10^6$ particles. However, we note that the plateau starts to bend and does not reproduce a constant density any more That we do not reach the target density even in the highest resolution is due to the fact that at the same time it is not possible to generate a perfect glass distribution and the correct model density. It is always a trade-of between the quality of the glass and the correct model density.}
	    	\label{fig:tophatdifferentresolutions}
	    \end{figure}
	    To investigate convergence on the algorithm with particle number, we plot the density for the plateau test case with $10^4$, $10^5$ and $10^6$ particles next to each other in Fig. \ref{fig:tophatdifferentresolutions}, zooming onto the top plateau. In all three runs we adjust the parameters such, that the initial degree of particle movement is of the same order to have a fair comparison. While the random sampling produces very similar results we see that the particle distribution develops differently in all three runs. The lower resolution runs converge much quicker than the higher resolution, as the mean particle distance is larger in this case. The plateau approaches the imposed model with increasing resolution, but shows more curvature as observed before. We can distinguish effects of the whole algorithm and of the inherent kernel smoothing. The latter is most apparent in the lowest resolution, where we see stronger curvature at the position of kinks in the density model. Finally, the scatter around the mean in the particle distribution is of the same order in all three runs.\\
	    \begin{figure}
	    	\centering
	    	\includegraphics[width=\columnwidth]{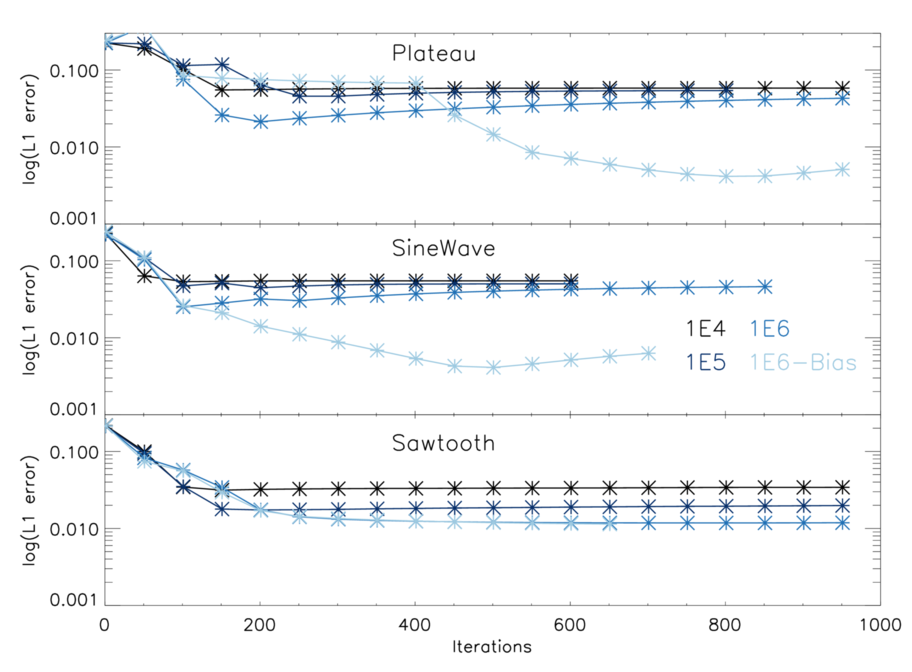}
	    	\caption{We show the L1 error as a function of the number of iterations of the WVT-algorithm for the top hat profile with gradients (top), the sine wave (middle) and the saw tooth density distribution (bottom). In the beginning the L1 error drops as a function of the iteration count. For the top hat profile with gradients the and the sin wave the L1 error increases again and saturates at few times $0.01$.  For the saw tooth the L1 drops and saturates at  $0.01$.}
	    	\label{fig:l1errorevolution}
	    \end{figure}
	    To investigate the behaviour of convergence with iterations and changing amount of particles quantitatively, we present the L1-error of all of our three standard tests in three different resolution steps in Fig. \ref{fig:l1errorevolution}. The L1 error is defined in e.g. \cite{Hopkins2015a} as
	    \begin{equation}
		    \mathrm{L1} = \frac{\sum_i \left| \rho_i - \rho_m \right|}{N_p}
	    \end{equation}
	    All tests have been carried out with a similar degree of movement in the initial setup, but without the redistribution algorithm. Since the L1 error is a global measurement, we only see the globally most prominent evolution in the beginning when the distribution quickly approaches the imposed density model. After about 100 steps, we observe hardly any change. Nevertheless, we observe a clear decrease of the L1 error with increasing particle number. According to the argumentation of \cite{Hopkins2015a}, the L1-error can not become better than $10^{-2} - 10^{-3}$ in classical SPH due to the $E_0$ gradient error of particle disorder \citep{Read2010}. This poses a hard lower bound on convergence with particle number.
	    
\subsection{Artificial bias correction}
		\label{sec:artificialbiascorrection}
\begin{figure}
	    	\centering
	    	\includegraphics[width=\columnwidth]{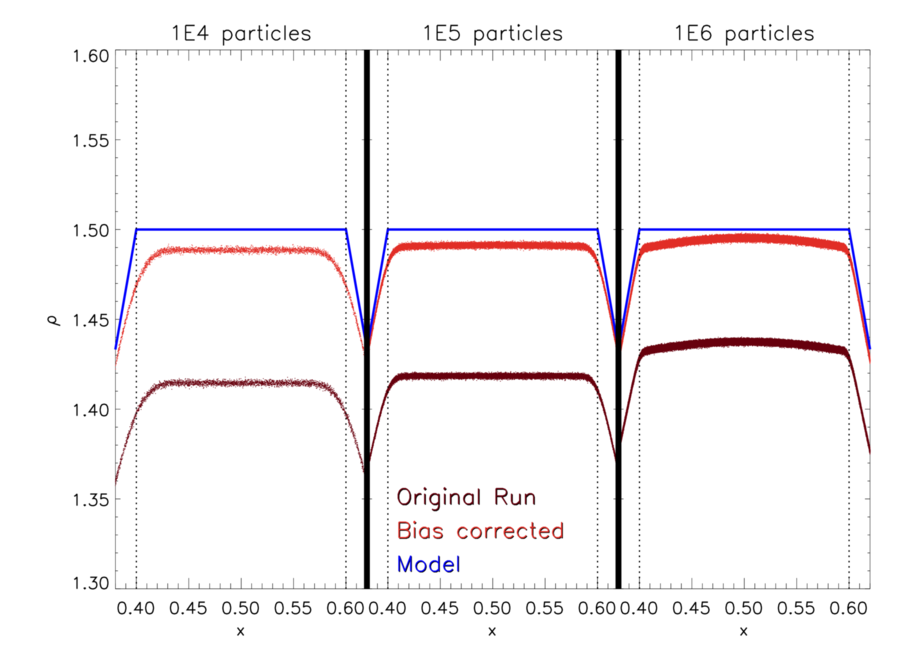}
	    	\caption{Comparison between the original best converged simulation and a run with employed bias correction. Over all resolutions the bias correction works well and fits the model density better, while still preserving the original shape.}
	    	\label{fig:biascorrected}
	    \end{figure}

To account for the over- and undershooting of the resulting density we employ an optional bias correction. 
For that we calculate the density deviation following the equation:
\begin{equation}
	\delta = \frac{\rho - \rho_m}{\rho_m - \bar{\rho} + \frac{\rho_m - \bar{\rho}}{\vert \rho_m - \bar{\rho} \vert} \bar{\rho}}, 
	\label{eq:artificialbias}
\end{equation}
with $\rho$, $\rho_m$ and $\bar{\rho}$ being achieved density, model density and mean density respectively. 
The resulting bias is calculated at the end of the run and prompted to the user.
This value can be passed in the parameter-file for the next run where the model density is then increased or decreased in the areas where we over- and undershoot:
\begin{equation}
	\rho = \rho_m + (\rho_m - \bar{\rho}) \cdot \delta,
\end{equation}
where $\delta$ denotes the bias factor from Eq. \ref{eq:artificialbias}.
This approximates the target density significantly more accurate, as can be seen in Fig. \ref{fig:biascorrected} where the plateau test was rerun with the bias correction switched on. Likewise, with this bias correction in place the L1 error (Fig. \ref{fig:l1errorevolution}) drops almost one order of magnitude and thereby approaches the theoretical hard lower bound of $10^{-2} - 10^{-3}$ referenced in the previous section.

\section{Applications}
	\label{sec:application}
	
	In this section we demonstrate typical use cases in astrophysics from standard code tests to studies of astrophysical objects.
	
	\subsection{Code testing: (Magneto-) Hydrodynamics}
		\label{sec:mhdtesting}

\begin{figure}
			\centering
			\includegraphics[width=\columnwidth]{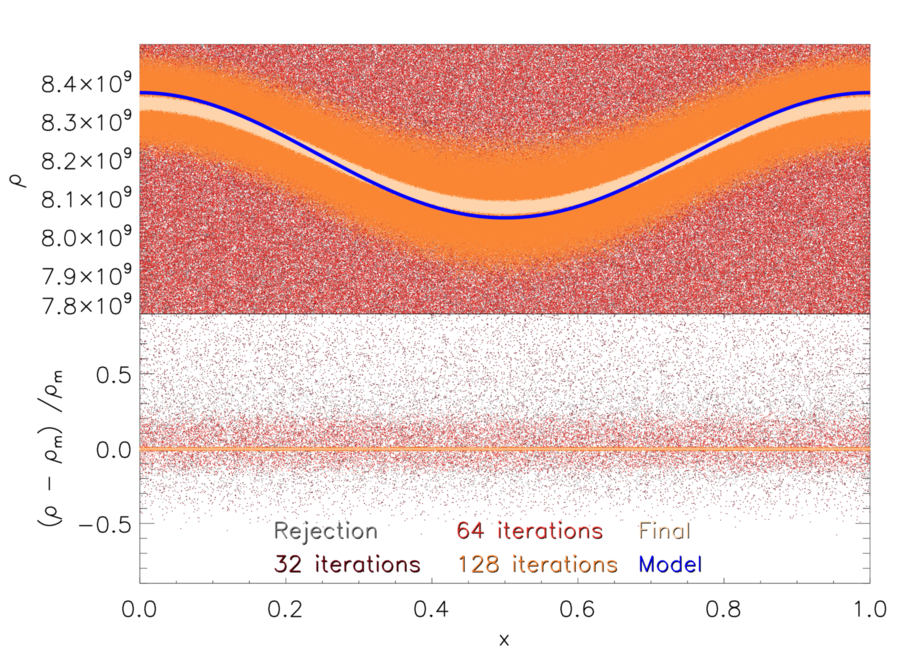}
			\caption{Zeldovich pancake initial conditions with $10^6$ particles. Again we start with a rejection sampling supported random distribution and plot the resulting density after several iteration steps. The result is very close to the analytical model.}
			\label{fig:zeldovichpancake}
		\end{figure}

		\begin{table}
			\centering
			\begin{tabular}{lll}
				\hline
				\multicolumn{2}{c}{Classical Hydrodynamic Tests}\\\hline
				Sod-Shock tube               	       & \cite{Sod1978}    \\
				Sedov-Blastwave			       & \cite{Hu2016}           \\
				Kelvin-Helmholtz Instability             & \cite{Mcnally2012}                \\
				Keplerian Ring          		       & \cite{Maddison1996}               \\
				Cold Blob               		       & \cite{Agertz2007}               \\
				Hydrostatic Sphere                       & \cite{Springel2005}               \\
				Evrard Collapse                          & \cite{Evrard1988}                \\
				Zeldovich Pancake			       & \cite{Zeldovich1970}                       \\
				Box-Test				       & \cite{Hess2010}                      \\
				Gresho-Vortex			       & \cite{Gresho1990}                      \\
				\hline
				\multicolumn{2}{c}{Classical Magnetohydrodynamic Tests}\\\hline
				Ryu-Jones Shock tube                     & \cite{Ryu1995b}               \\
				Fast-Rotor                               & \cite{Balsara1999}           \\
				Strong-Blast                             & \cite{Hopkins2016a}                \\
				Orszang-Tang-Vortex                      & \cite{Hopkins2016a}               \\
				Linear Alfven Wave                       & \cite{Stone2008}               \\
				Rayleigh-Taylor-Instability              & \cite{Abel2011}               \\
			\end{tabular}
			\caption{Hydrodynamic and Magnetohydrodynamic test cases implemented in the Code.}
			\label{tab:testproblems}
		\end{table}
		Our implementation can be used to provide high quality initial conditions for standard (magneto-) hydrodynamical benchmark tests (\cite{Springel2005}, \cite{Dolag2009}, \cite{Hu2014}, \cite{Beck2016}, \cite{Springel2010a}, \cite{Hopkins2015a} \cite{Hopkins2016a}, \cite{Hubber2017}). In particular, the WVT technique is useful to generate a well defined initial particle distribution with a low degree of noise. Since the method we present in this paper is capable of sampling arbitrary density profiles, we provide implementations of common benchmark tests listed in table \ref{tab:testproblems}, alongside references for the implemented models. \\
		Here we demonstrate a more complex example the Zeldovich Pancake \citep{Zeldovich1970}. We use the same initial configuration, which has been presented in \cite{Beck2016}:
		\begin{align}
			x(q,z) = q - \frac{1+z_{c}}{1+z} \frac{\sin(kq)}{k},
		\end{align} 
		with the wave number $k=2\pi /\lambda$ and the unperturbed coordinate $q$, as well as the collapse redshift $z_{c}=1$. The density profile is then given via
		\begin{align}
			\rho(x,z) = \frac{\rho_{0}}{1- \frac{1+z_{c}}{1+z} \cos(kq)},
		\end{align}
		with the initial density $\rho_{0}$. We use a Box with $64$ Mpc side length and sample the density profile with $10^6$ gas particles. The Zeldovich-Pancake is an important test problem for both numerical hydrodynamics and numerical integration in a comoving frame where time $t$ is replaced with the Hubble-function $H(t)$. The resulting initial conditions are shown in Fig. \ref{fig:zeldovichpancake}. The quality of the initial conditions are very well in line with our previous results.
		
	\subsection{Binary star formation}
		\label{sec:binarystarformation}
		
\begin{figure}
                \centering
                \includegraphics[width=\columnwidth]{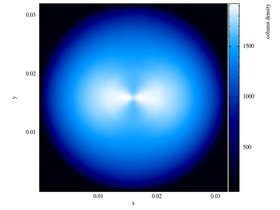}
                \caption{Integrated density of the Boss-Bodenheimer test case for binary star formation in the x-y-plane with $10^6$ particles after convergence. The result resembles the analytic model very well.}
                \label{fig:bossbodenheimer}
        \end{figure}

		In this section, we present the initial conditions for the Boss and Bodenheimer test case \citep{Boss1979} to simulate the formation of a binary system of stars. We follow the implementation presented in \cite{Springel2005} with the density profile
		\begin{align}
			\rho = \rho_{0}[1+0.1\cos(2\varphi)],
			\label{eq:bossbodenheimer}
		\end{align}
		using a central density $\rho_{0} = 3.82 \cdot 10^{-18}$ $\mathrm{g}/\mathrm{cm}^3$. Furthermore, we set up a sphere of radius $R=5.16 \cdot 10^{16}$ cm, a total mass of $1\mathrm{M}_{\odot}$, speed of sound $c_{s}=1.66 \cdot 10^{4}$ cm/s and a solid body rotation of $\omega = 7.2 \cdot 10^{-13}$ 1/s, as of \cite{Burkert1993}. In Fig. \ref{fig:bossbodenheimer} we show the rendered density profile in x-y plane with the perturbation in direction of $\varphi$.
		
	\subsection{Isolated galaxy cluster}
		\label{sec:isolatedcluster}
	
		The baryonic content of clusters of galaxies, in general, follows a beta-model profile \citep[e.g.][]{Cavaliere1978}, given by 
		\begin{align}
			\rho_{gas}(r) &= \rho_{0} \left[ 1 + \left( r / r_c \right)^2 \right]^{-3 \beta / 2} \quad , \label{eq:betamodel}
		\end{align}
		where the density within the core radius $r_c$ is constant and decays as a power-law outside the core. \citet{Donnert2014} set up initial conditions for idealised binary merger simulations in the now open source code \textsc{Toycluster}\footnote{https://github.com(jdonnert/toycluster)}, further improved by \citet{Donnert2017} by adding WVT relaxation. For simplicity, here we ignore the dark matter halo of the cluster as well as the temperature structure. The `cluster' is set up with general parameters $\rho_0 = 10^{-26}$~g/cm$^{-3}$, $r_c=20$~kpc, and $\beta=2/3$ which could resemble a cool-core cluster of galaxies. We sample the density profile using $10^6$ SPH particles.\\
		\begin{figure}
			\centering
			\includegraphics[width=\columnwidth]{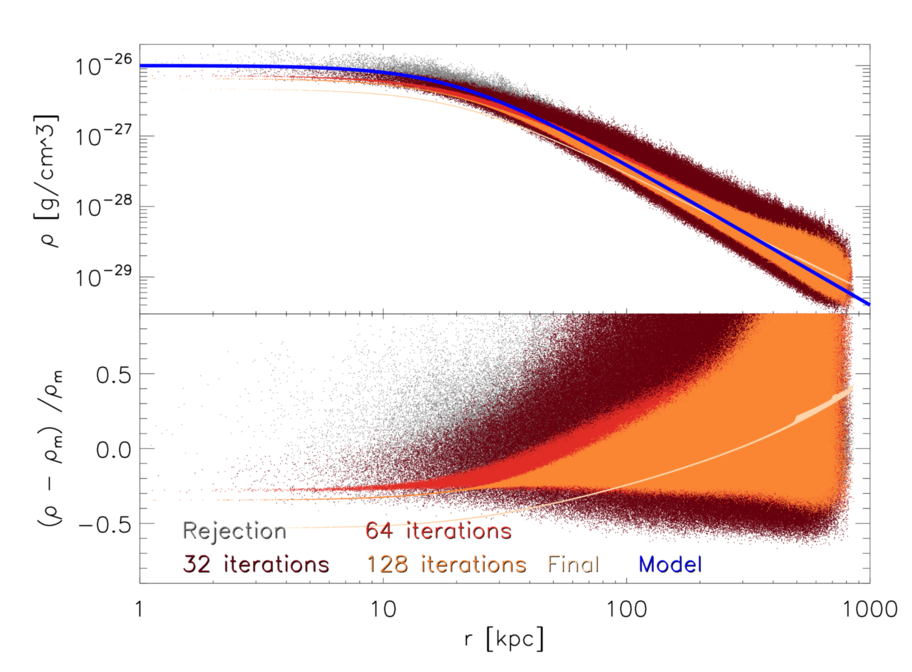}
			\caption{Beta profile for an isolated galaxy cluster with $10^6$ particles after the same iterations steps as previous tests. Due to the large density range these initial conditions are very challenging for our code and it takes more iterations for good convergence that with previous tests. Nevertheless, the result is very close to the desired model.}
			\label{fig:isolatedcluster}
		\end{figure}		
		We show the resulting density in Fig. \ref{fig:isolatedcluster} at $32$, $64$, $128$ and $1024$ iterations, where we hit the maximum number of iterations and the algorithm terminated. This happens due to the large range of more than three orders of magnitudes in density which we try to sample with only $10^6$ particles. In the low density region we can not properly resolve the gradient any more because the SPH kernels become very large, leading to a locally large spread in the density distribution early on which it relaxes away nicely. Furthermore, this results in quite big density errors in the low density regions of the resulting distribution. Judging these errors relatively to the mean density they are, however, quite small. Consequently, since the particle masses are calculated from the mean density, this over-density directly leads to an under-density in the central regime. Nevertheless our results are consistent with \cite{Donnert2014}, who successfully uses similar initial conditions for actual simulations of galaxy cluster mergers.
		
	\subsection{Reading in an initial setup: Image processing}
		\label{sec:imagereading}
		
		\begin{figure}
			\centering
			\includegraphics[width=\columnwidth]{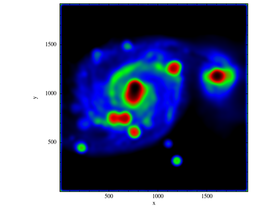}
			\caption{Example for initial conditions sampled according to a read image of the whirlpool galaxy M51a and it's companion M51b with $10^6$ particles.}
			\label{fig:m51}
		\end{figure}
		Finally, we present a special feature of our application which allows us to read in a given set of gridded data for the definition of a model density, instead of providing a hard coded analytic solution. An example where we directly used the brightness scale of an image utilising our built in PNG reading routine as proxy for the density, is shown in Fig. \ref{fig:m51}. The goal of astrophysical simulations is always the comparison to observations in order to learn more about the processes in the universe. Mostly only statistical comparisons can be made because it is extremely difficult to simulate an observed object exactly in all it's details. This feature might actually help here, taking in an observational image in order to model the underlying matter distribution. The implementation is currently being expanded to take in proper 3D data which might be for example de-projected from observational images. Other applications outside of astrophysics can be found, where one might for example build a CAD model of a 3D distribution \cite[see e.g.][]{Dauch2017}.
		
\section{Code usage}
	\label{sec:usage}
	
	In this final section we talk briefly about the actual usage of our application. It can be downloaded from github \citep{Donnert2017a} and will be made publicly available under the MIT license with the release of this paper. The current code version described here is v2.3. This section is split into two parts: First, we briefly state the compilation requirements. Second, we show the currently required parameters and describe how to actually run the application successfully.
	
	\subsection{Requirements and compilation}
		\label{sec:compilation}
		
		The code is written in OpenMP parallel C with the C99 standard and requires the gsl and gslcblas libraries. If the PNG reader shall be used, libpng is necessary in addition. There are no special requirements regarding the choice of compiler or any known issues with optimisation flags.\\
		\begin{table*}
			\centering
			\begin{tabular}{llp{8.5cm}}
				Option&Default&Description\\
				\hline
				SAVE\_WVT\_STEPS&ON&Write out a snapshot after each relaxation iteration\\
				SPH\_CUBIC\_SPLINE&OFF&Switch from Wendland C6 kernel to cubic spline\\
 				SPH\_WC2&OFF&Switch from Wendland C6 kernel to Wendland C2\\
 				REJECTION\_SAMPLING&ON&Use rejection sampling for initial particle distribution\\
 				PEANO\_SAMPLING&OFF&Use a Peano curve for the initial particle distribution\\
 				EAT\_PNG&ON&Compile with png reading support\\
 				TWO\_DIM&OFF&Switch from 3D to 2D, ignoring z coordinates\\
 				BRUTE\_FORCE\_NGB&OFF&Instead of a tree use a brute force neighbour finder (mainly used for debugging purposes)\\
 				OUTPUT\_DIAGNOSTICS&ON&Output additional diagnostics to file each iteration
			\end{tabular}
			\caption{Compilation options with default values and description.}
			\label{tab:compileoptions}
		\end{table*}
		For compilation we provide two identical approaches, a direct Makefile or, as required by some IDEs, a CMake file which then generates an own Makefile. Both approaches provide the possibility to adjust the compilation by setting some options which are translated into preprocessor defines. An overview is given in table \ref{tab:compileoptions}.
				
	\subsection{Parameters and running the application}
		\label{sec:runtime}
		
		\begin{table*}
			\centering
			\begin{tabular}{llp{8.5cm}}
				Parameter&Default&Description\\
				\hline
				Npart&100000&Amount of particles to distribute\\
				Maxiter&512&Maximum iterations of relaxation\\
				BiasCorrection&0.0&Correction factor to scale density model around a mean value\\
				MpsFraction&1.0&Scaling parameter for the particle displacements\\
				StepReduction&0.95&Factor to reduce step sizes in case of overshooting\\
				LimitMps&-1&Abort if particles moved farther than the local mean particle separation below this percentage\\
				LimitMps10&-1&Abort if particles moved farther than one tenth of the local mean particle separation below this percentage\\
				LimitMps100&-1&Abort if particles moved farther than one hundredth of the  the local mean particle separation below this percentage\\
				LimitMps1000&-1&Abort if particles moved farther than one thousandth of the  local mean particle separation below this percentage\\
				MoveFractionMin&0.001&Fraction of particles to redistribute maximally after the decay\\
				MoveFractionMax&0.01&Fraction of particles to redistribute maximally initially\\
				ProbesFraction&0.1&Fraction of particles to be probed for redistribution initially\\
				RedistributionFrequency&5&How many relaxation steps come between two steps of redistribution\\
				LastMoveStep&256&Exponential decay of the redistribution amount until this iteration, afterwards shut off\\
				Problem\_Flag&0&Main flag to chose the initial conditions model (full list in ics.par)\\
				Problem\_Subflag&0&Sub flag to chose the initial conditions model (full list in ics.par)\\
				PNG\_Filename&pic.png&Which PNG to read in, if the respective Flag and Subflag have been chosen
			\end{tabular}
			\caption{Runtime parameters with example values and descriptions.}
			\label{tab:parameters}
		\end{table*}
		In order to run the application, one has to provide a parameter file by path. The main code directory contains already a sample parameter file called \quotes{ics.par}. It lists all provided runtime configuration options to adjust the main algorithms of the code and chose the respective initial conditions model. A list of parameters including standard values and short explanations are given in table \ref{tab:parameters}.\\
		Besides these parameters one can change the behaviour of the OpenMP parallelisation by setting the usual environment variables to appropriate values, like\\OMP\_NUM\_THREADS.\\
		The output is written in standard GADGET type 2 binary format.
    
\section{Summary and Conclusions}
	\label{sec:conclusions}
	
	In this paper we presented a novel application to generate arbitrary initial conditions for SPH with a glass-like particle structure. The applied algorithms are based on a combination of weighted Voronoi Tesselation \citep{Diehl2012}, density aware random sampling \citep{VonNeumann1951,Peano1890} and an additional reshuffling of particles \citep{Metropolis1953}. We showed an overview of common methods to solve this task and illustrated the challenges that come along. We ran several test problems with different density features, like linear and non-linear gradients, kinks and jumps, and investigated thoroughly how our implementation performs when confronted with them. We found that our algorithms converge reasonably well towards the given density models while also iterating towards a glass structure, which we analysed in 3D using a radial distance distribution function. We showed, that converging towards the correct density and a glass distribution at the same time is extremely difficult, since both tasks slightly contradict each other. Particle have to be pushed apart from each other in order to remove clumps and holes in the particle distribution, but this results in particles being moved out of high density regions resulting in lower density maxima and consequently also higher minima. Therefore, one has to settle for an application dependent trade-off. We also analysed the convergence of our algorithms with variable particle number and see desired behaviour.\\
	Our application also inhibits a suite of various sets of initial conditions, including many standard test problems for (magneto-) hydrodynamics and the ability to produce idealised setups for a galaxy cluster \citep[as in][]{Donnert2014}. These tests are frequently required for code development of SPH codes and often the ability to judge the quality of a code was limited by pour initial conditions.\\
	We also presented a special feature which allows the code to sample a density model according to a PNG image instead of an analytic model. This might be useful to bridge observations and simulations further, since it can lead to the ability to simulate observed objects better. For that more development has to be done, though.\\
	Finally, we also presented the actual usage of our application, including a short description of all options and parameters which can be set at the moment (v2.1). The source code  will be made publicly available with the release of this paper. Further development of the project, may include support for more read in / write out formats, extending the possible choices of SPH kernels, expanding the 2D image reader eventually to 3D and even improvements on the core algorithms themselves. WVTICs is a flexible and versatile tool applicable for all kinds of problems. We welcome feedback and contributions via github.
    
\section*{Acknowledgements}

	The authors would like to thank their colleague's Martin Zintl and Stefan Heigl for useful discussions.
	UPS is funded  by the  Deutsche  Forschungsgemeinschaft  (DFG,  German 
Research Foundation) with the project number MO 2979/1-1.
	This product includes software developed by Greg Roelofs and contributors for the book, \quotes{PNG: The Definitive Guide,} published by O'Reilly and Associated.

\bibliography{sources}

\begin{thebibliography}{}

\bibitem[Abel, 2011]{Abel2011}
Abel, T. (2011).
\newblock {rpSPH: A novel smoothed particle hydrodynamics algorithm}.
\newblock {\em Mon. Not. R. Astron. Soc.}, 413(1):271--285.

\bibitem[Agertz et~al., 2007]{Agertz2007}
Agertz, O., Moore, B., Stadel, J., Potter, D., Miniati, F., Read, J., Mayer,
  L., Gawryszczak, A., Kravtsov, A., Nordlund, {\AA}., Pearce, F., Quilis, V.,
  Rudd, D., Springel, V., Stone, J.~M., Tasker, E., Teyssier, R., Wadsley, J.,
  and Walder, R. (2007).
\newblock {Fundamental differences between SPH and grid methods}.
\newblock {\em Mon. Not. R. Astron. Soc.}, 978:963--978.

\bibitem[Balsara and Spicer, 1999]{Balsara1999}
Balsara, D.~S. and Spicer, D.~S. (1999).
\newblock {A Staggered Mesh Algorithm Using High Order Godunov Fluxes to Ensure
  Solenoidal Magnetic Fields in Magnetohydrodynamic Simulations}.
\newblock {\em J. Comput. Phys.}, 149(2):270--292.

\bibitem[Bate et~al., 2013]{Bate2013}
Bate, M.~R., Tricco, T.~S., and Price, D.~J. (2013).
\newblock {Collapse of a molecular cloud core to stellar densities:
  Stellar-core and outflow formation in radiation magnetohydrodynamic
  simulations}.
\newblock {\em Mon. Not. R. Astron. Soc.}, 437(1):77--95.

\bibitem[Beck et~al., 2016]{Beck2016}
Beck, A.~M., Murante, G., Arth, A., Remus, R.-S., Teklu, A.~F., Donnert, J.
  M.~F., Planelles, S., Beck, M.~C., F{\"{o}}rster, P., Imgrund, M., Dolag, K.,
  and Borgani, S. (2016).
\newblock {An improved SPH scheme for cosmological simulations}.
\newblock {\em Mon. Not. R. Astron. Soc.}, 455:2110--2130.

\bibitem[Bonafede et~al., 2011]{Bonafede2011}
Bonafede, A., Dolag, K., Stasyszyn, F., Murante, G., and Borgani, S. (2011).
\newblock {A non-ideal magnetohydrodynamic gadget: simulating massive galaxy
  clusters}.
\newblock {\em Mon. Not. R. Astron. Soc.}, 418(4):2234--2250.

\bibitem[Boss and Bodenheimer, 1979]{Boss1979}
Boss, A.~P. and Bodenheimer, P. (1979).
\newblock {Fragmentation in a rotating protostar - A comparison of two
  three-dimensional computer codes}.
\newblock {\em ApJ}, 234:289.

\bibitem[Burkert and Bodenheimer, 1993]{Burkert1993}
Burkert, A. and Bodenheimer, P. (1993).
\newblock {Multiple fragmentation in collapsing protostars}.
\newblock {\em Mon. Not. R. Astron. Soc.}, 264.

\bibitem[Cavaliere and Fusco-Femiano, 1978]{Cavaliere1978}
Cavaliere, A. and Fusco-Femiano, R. (1978).
\newblock {The Distribution of Hot Gas in Clusters of Galaxies}.
\newblock {\em Astron. Astrophys.}, 70:677.

\bibitem[Dauch et~al., 2017]{Dauch2017}
Dauch, T.~F., Okraschevski, M., Keller, M.~C., Braun, S., Wieth, L.,
  Chaussonnet, G., Koch, R., and Bauer, H. (2017).
\newblock {Preprocessing Workflow for the Initialization of SPH Predictions
  based on Arbitrary CAD Models}.
\newblock In {\em Proc. SPHERIC 2017 – 12th Int. SPHERIC Work.}, Ourense.

\bibitem[Dehnen and Aly, 2012]{Dehnen2012}
Dehnen, W. and Aly, H. (2012).
\newblock {Improving convergence in smoothed particle hydrodynamics simulations
  without pairing instability}.
\newblock {\em Mon. Not. R. Astron. Soc.}, 1082.

\bibitem[Diehl et~al., 2012]{Diehl2012}
Diehl, S., Rockefeller, G., Fryer, C.~L., Riethmiller, D., and Statler, T.~S.
  (2012).
\newblock {Generating Optimal Initial Conditions for Smooth Particle
  Hydrodynamics Simulations}.
\newblock {\em Mon. Rev.}, 000(November):30.

\bibitem[Dolag and Stasyszyn, 2009]{Dolag2009}
Dolag, K. and Stasyszyn, F. (2009).
\newblock {An MHD gadget for cosmological simulations}.
\newblock {\em Mon. Not. R. Astron. Soc.}, 398(4):1678--1697.

\bibitem[Donnert et~al., 2017a]{Donnert2017a}
Donnert, J., Arth, A., Halbesma, T., and Steinwandel, U. (2017a).
\newblock {GIthub repository for WVTICs}.

\bibitem[Donnert, 2014]{Donnert2014}
Donnert, J. M.~F. (2014).
\newblock {Initial conditions for idealized clusters mergers, simulating 'El
  Gordo'}.
\newblock {\em Mon. Not. R. Astron. Soc.}, 438(3):1971--1984.

\bibitem[Donnert et~al., 2017b]{Donnert2017}
Donnert, J. M.~F., Beck, A.~M., Dolag, K., and R{\"{o}}ttgering, H. J.~A.
  (2017b).
\newblock {Simulations of the Galaxy Cluster CIZA J2242.8+5301 I: Thermal Model
  and Shock Properties}.
\newblock 19(March):1--19.

\bibitem[Evrard, 1988]{Evrard1988}
Evrard, A.~E. (1988).
\newblock {Beyond N-body: 3D cosmological gas dynamics}.
\newblock {\em Mon. Not. R. Astron. Soc.}, 235.

\bibitem[Frenkel and Smit, 2001]{Frenkel2001}
Frenkel, D. and Smit, B. (2001).
\newblock {\em {Understanding Molecular Simulation: From Algorithms to
  Applications}}.
\newblock Computational science series. Elsevier Science.

\bibitem[Gaburov and Nitadori, 2011]{Gaburov2011}
Gaburov, E. and Nitadori, K. (2011).
\newblock {Astrophysical weighted particle magnetohydrodynamics}.
\newblock {\em Mon. Not. R. Astron. Soc.}, 414(1):129--154.

\bibitem[Gingold and Monaghan, 1977]{Gingold1977}
Gingold, R.~a. and Monaghan, J.~J. (1977).
\newblock {Smoothed particle hydrodynamics-theory and application to
  non-spherical stars}.
\newblock {\em Mon. Not. R. Astron. Soc.}, 181:375--389.

\bibitem[Gresho and Chan, 1990]{Gresho1990}
Gresho, P.~M. and Chan, S.~T. (1990).
\newblock {On the theory of semi‐implicit projection methods for viscous
  incompressible flow and its implementation via a finite element method that
  also introduces a nearly consistent mass matrix. Part 2: Implementation}.
\newblock {\em Int. J. Numer. Methods Fluids}, 11(5):621--659.

\bibitem[Hansen and Verlet, 1969]{Hansen1969}
Hansen, J.-P. and Verlet, L. (1969).
\newblock {Phase Transitions of the Lennard-Jones System}.
\newblock {\em Phys. Rev.}, 184(1):151--161.

\bibitem[He{\ss} and Springel, 2010]{Hess2010}
He{\ss}, S. and Springel, V. (2010).
\newblock {Particle hydrodynamics with tessellation techniques}.
\newblock {\em Mon. Not. R. Astron. Soc.}, 406(4):2289--2311.

\bibitem[Hirschmann et~al., 2014]{Hirschmann2014}
Hirschmann, M., Dolag, K., Saro, A., Bachmann, L., Borgani, S., and Burkert, A.
  (2014).
\newblock {Cosmological simulations of black hole growth: AGN luminosities and
  downsizing}.
\newblock {\em Mon. Not. R. Astron. Soc.}, 442(3):2304--2324.

\bibitem[Hopkins, 2015]{Hopkins2015a}
Hopkins, P.~F. (2015).
\newblock {A new class of accurate, mesh-free hydrodynamic simulation methods}.
\newblock {\em Mon. Not. R. Astron. Soc.}, 450:53--110.

\bibitem[Hopkins and Raives, 2016]{Hopkins2016a}
Hopkins, P.~F. and Raives, M.~J. (2016).
\newblock {Accurate, meshless methods for magnetohydrodynamics}.
\newblock {\em Mon. Not. R. Astron. Soc.}, 455(1):51--88.

\bibitem[Hu et~al., 2016]{Hu2016}
Hu, C.-y., Naab, T., Walch, S., Glover, S. C.~O., and Clark, P.~C. (2016).
\newblock {Star formation and molecular hydrogen in dwarf galaxies: a
  non-equilibrium view}.
\newblock {\em Mon. Not. R. Astron. Soc.}, 458(4):3528--3553.

\bibitem[Hu et~al., 2014]{Hu2014}
Hu, C.-Y., Naab, T., Walch, S., and Moster, B.~P. (2014).
\newblock {SPHGal : Smoothed Particle Hydrodynamics with improved accuracy for
  galaxy simulations}.

\bibitem[Hubber and Rosotti, 2016]{Hubber2016}
Hubber, D. and Rosotti (2016).
\newblock {GANDALF: Graphical Astrophysics code for N-body Dynamics And
  Lagrangian Fluids}.

\bibitem[Hubber et~al., 2017]{Hubber2017}
Hubber, D.~A., Rosotti, G.~P., and Booth, R.~A. (2017).
\newblock {GANDALF - Graphical Astrophysics code for N-body Dynamics And
  Lagrangian Fluids}.
\newblock 33(September):1--33.

\bibitem[Kotarba et~al., 2011]{Kotarba2011}
Kotarba, H., Lesch, H., Dolag, K., Naab, T., Johansson, P.~H., Donnert, J., and
  Stasyszyn, F.~A. (2011).
\newblock {Galactic m{\'{e}}nage {\`{a}} trois: Simulating magnetic fields in
  colliding galaxies}.
\newblock {\em Mon. Not. R. Astron. Soc.}, 415(4):3189--3218.

\bibitem[Liska and Wendroff, 2003]{Liska2003}
Liska, R. and Wendroff, B. (2003).
\newblock Comparison of several difference schemes on 1d and 2d test problems
  for the euler equations.
\newblock {\em SIAM Journal on Scientific Computing}, 25(3):995--1017.

\bibitem[Lucy, 1977]{Lucy1977}
Lucy, L. (1977).
\newblock {A numerical approach to the testing of the fission hypothesis}.
\newblock {\em Astron. J.}, 82(12):1013--1024.

\bibitem[Maddison et~al., 1996]{Maddison1996}
Maddison, S.~T., Murray, J.~R., and Monaghan, J.~J. (1996).
\newblock {SPH Simulations of Accretion Disks and Narrow Rings}.
\newblock {\em Publ. Astron. Soc. Aust.}, 13.

\bibitem[Mcnally et~al., 2012]{Mcnally2012}
Mcnally, C.~P., Lyra, W., and Passy, J.-c. (2012).
\newblock {a Well-Posed Kelvin – Helmholtz Instability Test and Comparison}.
\newblock 18.

\bibitem[Metropolis et~al., 1953]{Metropolis1953}
Metropolis, N., Rosenbluth, A.~W., Rosenbluth, M.~N., Teller, A.~H., and
  Teller, E. (1953).
\newblock {Equation of state calculations by fast computing machines}.
\newblock {\em J. Chem. Phys.}, 21(6):1087--1092.

\bibitem[Monaghan and Price, 2006]{Monaghan2006}
Monaghan, J.~J. and Price, D.~J. (2006).
\newblock {Toy Stars in two dimensions}.
\newblock {\em Mon. Not. R. Astron. Soc.}, 365(3):991--1006.

\bibitem[Muder, 1988]{Muder1988}
Muder, D.~J. (1988).
\newblock {Putting the Best Face of a Voronoi Polyhedron}.
\newblock In {\em Proc. London Math. Soc.}, volume~56, pages 329--348.

\bibitem[Pakmor and Springel, 2013]{Pakmor2013}
Pakmor, R. and Springel, V. (2013).
\newblock {Simulations of magnetic fields in isolated disc galaxies}.
\newblock {\em Mon. Not. R. Astron. Soc.}, 432(1):176--193.

\bibitem[Peano, 1890]{Peano1890}
Peano, G. (1890).
\newblock {Sur une courbe, qui remplit toute une aire plane}.
\newblock {\em Math. Ann.}, 36(1):157--160.

\bibitem[Price, 2012]{Price2012}
Price, D.~J. (2012).
\newblock {Smoothed particle hydrodynamics and magnetohydrodynamics}.
\newblock {\em J. Comput. Phys.}, 231(3):759--794.

\bibitem[Price and Monaghan, 2005]{Price2005}
Price, D.~J. and Monaghan, J.~J. (2005).
\newblock {Smoothed particle magnetohydrodynamics - III. Multidimensional tests
  and the div B = 0 constraint}.
\newblock {\em Mon. Not. R. Astron. Soc.}, 364(2):384--406.

\bibitem[Read et~al., 2010]{Read2010}
Read, J.~I., Hayfield, T., and Agertz, O. (2010).
\newblock {Resolving mixing in smoothed particle hydrodynamics}.
\newblock {\em Mon. Not. R. Astron. Soc.}, 405(3):1513--1530.

\bibitem[Ryu et~al., 1995]{Ryu1995b}
Ryu, D., Jones, T.~W., and Frank, A. (1995).
\newblock {Numerical Magnetohydrodynamics in Astrophysics: Algorithm and Tests
  for Multidimensional Flow}.
\newblock {\em Astrophys. J.}, 452:758--796.

\bibitem[Schaye et~al., 2015]{Schaye2015}
Schaye, J., Crain, R.~A., Bower, R.~G., Furlong, M., Schaller, M., Theuns, T.,
  {Dalla Vecchia}, C., Frenk, C.~S., Mccarthy, I.~G., Helly, J.~C., Jenkins,
  A., Rosas-Guevara, Y.~M., White, S.~D., Baes, M., Booth, C.~M., Camps, P.,
  Navarro, J.~F., Qu, Y., Rahmati, A., Sawala, T., Thomas, P.~A., and Trayford,
  J. (2015).
\newblock {The EAGLE project: Simulating the evolution and assembly of galaxies
  and their environments}.
\newblock {\em Mon. Not. R. Astron. Soc.}, 446(1):521--554.

\bibitem[Sod, 1978]{Sod1978}
Sod, G.~A. (1978).
\newblock {A survey of several finite difference methods for systems of
  nonlinear hyperbolic conservation laws}.
\newblock {\em J. Comput. Phys.}, 27(1):1--31.

\bibitem[Springel, 2005]{Springel2005}
Springel, V. (2005).
\newblock {User guide for GADGET-2}.
\newblock pages 1--46.

\bibitem[Springel, 2010]{Springel2010a}
Springel, V. (2010).
\newblock {E pur si muove: Galilean-invariant cosmological hydrodynamical
  simulations on a moving mesh}.
\newblock {\em Mon. Not. R. Astron. Soc.}, 401(2):791--851.

\bibitem[{Steinwandel} et~al., 2019]{Steinwandel2019}
{Steinwandel}, U.~P., {Beck}, M.~C., {Arth}, A., {Dolag}, K., {Moster}, B.~P.,
  and {Nielaba}, P. (2019).
\newblock {Magnetic buoyancy in simulated galactic discs with a realistic
  circumgalactic medium}.
\newblock {\em Mon. Not. R. Astron. Soc.}, 483(1):1008--1028.

\bibitem[Stone et~al., 2008]{Stone2008}
Stone, J.~M., Gardiner, T.~a., Teuben, P., Hawley, J.~F., and Simon, J.~B.
  (2008).
\newblock {Athena: A New Code for Astrophysical MHD}.
\newblock {\em Astrophys. J. Suppl. Ser.}, 178(1):137--177.

\bibitem[Vogelsberger et~al., 2014]{Vogelsberger2014}
Vogelsberger, M., Genel, S., Springel, V., Torrey, P., Sijacki, D., Xu, D.,
  Snyder, G., Nelson, D., and Hernquist, L. (2014).
\newblock {Introducing the illustris project: Simulating the coevolution of
  dark and visible matter in the universe}.
\newblock {\em Mon. Not. R. Astron. Soc.}, 444(2):1518--1547.

\bibitem[{Von Neumann}, 1951]{VonNeumann1951}
{Von Neumann}, J. (1951).
\newblock {Various techniques used in connection with random digits. Monte
  Carlo methods}.
\newblock {\em Natl. Bur. Stand.}, 12:36--38.

\bibitem[White, 1994]{White1994}
White, S. D.~M. (1994).
\newblock {Formation and Evolution of Galaxies: Les Houches Lectures}.

\bibitem[Zeldovich, 1970]{Zeldovich1970}
Zeldovich, Y.~B. (1970).
\newblock {Gravitational Instability: An Approximate Theory for Large Density
  Perturbations}.
\newblock {\em Astron. Astrophys.}, 5:84--89.

\end{thebibliography}

\end{document}